\documentclass[12pt]{article}
\usepackage{qcircuit}
\usepackage{times}
\usepackage{braket}
\usepackage{amsmath}
\usepackage{mathtools}
\usepackage{graphicx}
\usepackage{caption}
\usepackage{subcaption}
\title{Two quantum algorithms for communication between spacelike separated locations}

\author
{Amitava Datta \\
Department of Computer Science and Software Engineering,\\ University of Western Australia,\\ 
35 Stirling Highway, Perth, WA6009, Australia.}

\begin{document}

\date{}

\maketitle
\begin{abstract}
The `no communication' theorem prohibits superluminal communication by showing that any measurement by Alice on an entangled system cannot change the reduced density matrix of Bob's state, and hence the expectation value of any measurement operator that Bob uses remains the same. We argue that the proof of the `no communication' theorem is incomplete and superluminal communication is possible through state discrimination in a higher-dimensional Hilbert space using ancilla qubits. 
We propose two quantum algorithms through state discrimantion for communication between two observers Alice and Bob, situated at spacelike separated locations. 
Alice and Bob share one qubit each of a Bell state $\frac{1}{\sqrt 2}(\ket{00}+\ket{11})$. While sending classical information, Alice measures her qubit and collapses the state of Bob's qubit in two different ways depending on whether she wants to send $0$ or $1$. Alice's first measurement is in the computational basis, and the second measurement is again in the computational basis after applying Hadamard transform to her qubit.
Bob's first algorithm detects the classical bit with probability of error $<\frac{1}{2^k}$, but Alice and Bob need to share $k$ Bell states for communicating a single classical bit. Bob's second algorithm is more complex, but Bob can detect the classical bit deterministically using four ancilla qubits. We also discuss possible applications of our algorithms. 
\end{abstract}

\flushbottom
\maketitle

\thispagestyle{empty}

\section{Introduction}
Nonlocality inherent in quantum mechanics was first pointed out by Einstein, Podolsky and Rosen~\cite{einstein} in their seminal 1935 paper. Their argument was that the apparent instantaneous effect of measuring one qubit of an entangled state on the other qubit in spacelike separated locations was an indication that quantum mechanics was an incomplete theory. However Bell~\cite{bell} in his 1965 paper showed that no deterministic hidden variable theory can make predictions that are consistent with quantum mechanics. This firmly established the inherent nonlocal nature of quantum mechanics. One natural 
question that arose after Bell's work is, whether this nonlocal nature of quantum mechanics can be utilized for sending 
information between spacelike seprated parties instantaneously. Several proposals were made to exploit the nonlocal 
correlations for superluminal signalling, or transfer of information~\cite{cufaro,popper,herbert,herbert1,greenberger}. However, all of these proposals either assumed 
operations that are not permitted by quantum mechanics, e.g., cloning of a quantum state, non-unitary transformations, or 
complex measurement processes that are impractical or not achievable with known technologies. The problems with these proposals were pointed out in several papers~\cite{araki,yanase,ghir,ghir1,ghir2,ghir3}. In particular Ghirardi~\cite{ghir1} summarized and refuted many of these proposals in a review article. The proposals and their refutations have resulted in two important {\em no go} theorems in quantum mechanics, the {\em no cloning} and {\em no communication} theorems. In particular, the 
{\em no communication} theorem was proven to show that the nonlocal correlations in EPR like scenarios cannot be used for 
sending information. The correletaions get nullified due to the probabilistic nature of the quantum measurement process.

Several proofs of the no communication theorem have been proposed over the years~\cite{shimony,ghir,bohm}, and all of them have an  implicit assumption that Bob can only try to get classical information from the expectation value of  a measurement operator. The first party Alice measures her qubits and the second party Bob then  measures his collapsed qubits to detect the changes in his collapsed states in a statistical sense, by collecting measurement statistics. The proofs of the no communication theorem 
show that the reduced density matrix of Bob remains the same irrespective of Alice's measurement. As a result, the expectation value of any measurement operator Bob uses is identical irrespective of Alice's measurement. We argue that these proofs of the `no communication' theorem do not rule out communication through state discrimination, when Bob identifies the single collapsed states due to Alice's measurements. 

The rest of the paper is organized as follows. We discuss communication through state discrimination in Section~\ref{discrimination}, we discuss the two algorithms in Section~\ref{algorithms}, and we conclude in Section~\ref{conclusion}.

\section{Communication through state discrimination}
\label{discrimination}

Quantum mechanics is a statistical theory as the measurement results can only give a probability distribution of possible outcomes. However, there are specific instances when a single measurement result on a state can give important information when a priori information is available about the measured state~\cite{ziman}. 
We argue that though the proof of the `no communication' theorem is correct in a statistical sense, it does not prevent communication through discrimination of quantum states. 

We reduce the communication problem to a state discrimination problem in the following way. Alice and Bob share one qubit each of a Bell state $\frac{1}{2}(\ket{00}+\ket{11})$. When Alice makes a measurement on her qubit, she can collapse the state of Bob's qubit in two different ways, in the first case by measuring in the computational basis and in the second case measuring in the computational basis after applying Hadamard transform to her qubit. Alice's first measurement in the computational basis collapses Bob's qubit to the set of states $S_1=\{\ket{0}, \ket{1}\}$, and Alice's second measurement in the computational basis after applying Hadamard transform to her qubit collapses Bob's qubit to the set of states $S_2=\{\frac{1}{2}(\ket{0}+\ket{1}), \frac{1}{2}(\ket{0}-\ket{1}\}$. Hence the result of Alice's measurement is the collapse of Bob's state to either a state from $S_1$ or a state from $S_2$. Note that Bob's reduced density matrix for these two sets are identical, and Bob cannot differentiate between these two sets statistically by using a local measurement operator. However, Bob knows these two possibilities a priori and Bob's job is to detect which set his state belongs to after each of Alice's measurements. If Bob detects his state to be in $S_1$, Alice is communicating the classical bit $0$, and if Bob's state is in $S_2$, Alice is communicating the classical bit $1$. We show that Bob can detect this either probabilistically or deterministically in two different algorithms, proving that it is possible for Alice and Bob to communicate even when they are at spacelike separated locations. We discuss these two algorithms in the next section and discuss the relevant work on state discrimination in this section. 

 The state discrimination problem is trivial if the two states $\ket{\psi_1}$ and $\ket{\psi_2}$ Alice prepares are orthogonal, as Bob can determine the state in a single measurement. Hence distinguishing between non-orthogonal states is the challenge, i.e.,when $\braket{\psi_1|\psi_2}\neq 0$. It is impossible to discriminate between non-orthogonal states in general~\cite{nielsen}. Chefles~\cite{chefles} showed that it is also impossible to distinguish between states that are linearly dependent. Ivanovic~\cite{ivanovic} was the first to consider this problem, and showed that it is possible to discriminate between non-orthogonal states if inconclusive results are allowed, in other words two non-orthogonal states cannot be discriminated perfectly. Ivanovic's procedure can be implemented using von Neumann measurements. If the measurement directions are $\ket{\psi_1}$ and $\ket{\psi_1^\perp}$, where $\ket{\psi_1^\perp}$ is orthogonal to $\ket{\psi_1}$, a success in measuring $\ket{\psi_1^\perp}$ in the $\{\ket{\psi_1},\ket{\psi_1^\perp}\}$ basis will indicate with certainty that the state is $\ket{\psi_2}$. However, a measurement result of ${\ket{\psi_1}}$ is inconclusive (the same argument can be repeated with a $\{\ket{\psi_2},\ket{\psi_2^\perp}\}$ basis. Ivanovic~\cite{ivanovic} showed that a sequence of measurements (potentially infinite in number) can improve the detection probability. 
Later Dieks~\cite{dieks} and Peres~\cite{peres1} showed that it is possible to implement Ivanovic's scheme with a single POVM (positive operator valued measurement), though the measurement can be inconclusive with some probability. Peres~\cite{peres1} showed that the probability of inconclusive measurement is $|\braket{\psi_1|\psi_2}|$, and this is optimal~\cite{hillery}. 

Neumark's theorem ~\cite{neumark,peres} states that every POVM can be realised as von Neumann projective measurements in a larger Hilbert space compared to the Hilbert space of the original system. These larger Hilbert spaces are usually constructed by using additional qubits called ancilla~\cite{peres}. Our problem is to distinguish between a state in $S_1$, and a state in $S_2$. The states in these two sets are linearly independent, but not orthogonal. 
Our approach is to use a two level ancilla system, we first increase the dimenionality of the system by entangling the original states in $S_1$ and $S_2$ with an ancilla qubit, using the CNOT gate. The aim is to make the systems in $S_2$ orthogonal to the systems in $S_1$. The systems in $S_2$ are now the two Bell states $\frac{1}{\sqrt 2}(\ket{00}+\ket{11})$ and $\frac{1}{\sqrt 2}(\ket{00}-\ket{11})$. Note that these two states are still not orthogonal to the states in $S_1$, $\ket{00}$ and $\ket{11}$ (after entangling with the first ancilla). However, applying the Hadamard transform to the two qubits of the state $\frac{1}{\sqrt 2}(\ket{00}-\ket{11})$, gives us the state $\frac{1}{\sqrt 2}(\ket{+-}+\ket{-+})$ in the Hadamard basis~\cite{ziman}, when $\ket{+}=\frac{1}{\sqrt 2}(\ket{0}+\ket{1})$ and $\ket{-}=\frac{1}{\sqrt 2}(\ket{0}-\ket{1})$. This can be shown either by direct calculation, by replacing $\ket{0}=\frac{1}{\sqrt 2}(\ket{+}+\ket{-})$ and $\ket{1}=\frac{1}{\sqrt 2}(\ket{+}-\ket{-})$, or by applying the $4$-dimensional Hadamard transform $H\otimes H=\frac{1}{\sqrt 2}\begin{pmatrix}1&1\\1&-1\end{pmatrix}\otimes \frac{1}{\sqrt 2}\begin{pmatrix}1&1\\1&-1\end{pmatrix}=\frac{1}{2}\begin{pmatrix}1&1&1&1\\1&-1&1&-1\\1&1&-1&-1\\1&-1&-1&1\end{pmatrix}$. Hence, $H^{\otimes2}\frac{1}{\sqrt 2}(\ket{00}-\ket{11})=\frac{1}{2}\begin{pmatrix}1&1&1&1\\1&-1&1&-1\\1&1&-1&-1\\1&-1&-1&1\end{pmatrix}\frac{1}{\sqrt 2}\begin{pmatrix}1\\0\\0\\-1\end{pmatrix}=\frac{1}{\sqrt 2}\begin{pmatrix}0\\1\\1\\0\end{pmatrix}$. This state is $\frac{1}{\sqrt 2}(\ket{+-}+\ket{-+})$ in the Hadamard basis. It can be seen easily that $H^{\otimes 2}\ket{00}=\ket{++}$ and $H^{\otimes 2}\ket{11}=\ket{--}$. Hence, the Hadamard transform of $\frac{1}{\sqrt 2}(\ket{00}-\ket{11})$ is orthogonal to the Hadamard transforms of the states in  $S_1$. This can be detected by using another ancilla qubit. Though $\frac{1}{\sqrt 2}(\ket{++}+\ket{--})$ is not orthogonal to the Hadamard transforms of the states in $S_1$, we can flip the relative phase of $\frac{1}{\sqrt 2}(\ket{0}+\ket{1})$ for converting it to the state $\frac{1}{\sqrt 2}(\ket{0}-\ket{1})$ and subsequently to the state $\frac{1}{\sqrt 2}(\ket{+-}+\ket{-+})$. We give the details in the next section.

\section{The algorithms}
\label{algorithms}

Our algorithms use entanglement for collapsing Bob's state in two different ways by Alice depending on the classical bit $0$ or $1$ she wants to send. Bob then identifies these different states by using ancilla qubits. Our algorithms have different qubit-circuit tradeoffs. Bob's first algorithm can detect the two classsical bits with probability of error $<\frac{1}{2^k}$ if Alice and Bob share $k$ Bell states for communicating a single classical bit, and Bob uses $k$ ancilla qubits. Bob can detect the classical bit deterministically with a more complex circuit and $4$ ancilla qubits in the second algorithm. As discussed before, our algorithms are based on discrimination of individual collapsed states at Bob's end.  The current lower bound of `spooky action at a distance', or speed of collapse of entanglement is four orders of magnitude of the speed of light~\cite{Yin}, hence our algorithms can be used for communicating between spacelike separated locations. 

\subsection{Alice's encoding algorithm}

We consider communication as a mechanism to send an ordered sequence of classical bits by the sender to the receiver. Alice wants to send a sequence of classical bits of her choice to the receiver Bob. Alice prepares maximally entangled Bell states $\ket{\Phi^+}=\frac{1}{\sqrt 2}(\ket{00}+\ket{11})$ for sending classical bits. Alice keeps the first qubit of $\ket{\Phi^+}$ and Bob takes the second qubit to a spacelike separated location. Alice measures her qubit in two different ways when she wants to send the classical bit $0$ or $1$. 

If Alice wants to send $0$, she measures her qubit in the computational basis. She gets the state $\ket{0}$ with probability $\frac{1}{2}$ and Bob's qubit is in the state $\psi_1=\ket{0}$. Alice gets the state $\ket{1}$ with probabilty $\frac{1}{2}$ and Bob's qubit is in the state $\psi_2=\ket{1}$. 

If Alice wants to send $1$, she first applies Hadamard transform to her qubit and then measures it in the computational basis. The result of applying Hadamard transform is: $\frac{1}{2}(\ket{0}+\ket{1})\ket{0}+\frac{1}{2}(\ket{0}-\ket{1})\ket{1}=\frac{1}{2}\ket{0}(\ket{0}+\ket{1})+\frac{1}{2}\ket{1}(\ket{0}-\ket{1})$. Hence Bob's state collapses to either $\ket{\psi_3}=\frac{1}{\sqrt 2}(\ket{0}+\ket{1})$, or 
$\ket{\psi_4}=\frac{1}{\sqrt 2}(\ket{0}-\ket{1})$ after Alice measures her qubit in the computational basis.

\subsection{Bob's decoding algorithms}

We use the single qubit Hadamard transform $H=\frac{1}{\sqrt 2}\begin{pmatrix}1&1\\1&-1\end{pmatrix}$, and two qubit Hadamard transform $H^{\otimes2}=H\otimes H=\frac{1}{2}\begin{pmatrix}1&1&1&1\\1&-1&1&-1\\1&1&-1&-1\\1&-1&-1&1\end{pmatrix}$. We also use the phase-flip or $Z$ transform (Pauli $\sigma_z$ operator) $\begin{pmatrix}1&0\\0&-1\end{pmatrix}$.  We need the two Bell states $\ket{\Phi^+}=\frac{1}{\sqrt 2}(\ket{00}+\ket{11})$, $\ket{\Phi^-}=\frac{1}{\sqrt 2}(\ket{00}-\ket{11})$. The single qubit basis states in the Hadamard basis are $\ket{+}=\frac{1}{\sqrt 2}(\ket{0}+\ket{1})$ and $\ket{-}=\frac{1}{\sqrt 2}(\ket{0}-\ket{1})$. Conversely, the computational basis states in the Hadamard basis are  $\ket{0} = \frac{1}{\sqrt 2}(\ket{+}+\ket{-})$ and $\ket{1}=\frac{1}{\sqrt 2}(\ket{+}-\ket{-})$. It can be verified easily by substituting for $\ket{0}$ and $\ket{1}$ that the Hadamard transforms of the two Bell states are: $H^{\otimes2}\ket{\Phi^+}=\frac{1}{\sqrt 2}(\ket{++}+\ket{--})$ and $H^{\otimes2}\ket{\Phi^-}=\frac{1}{\sqrt 2}(\ket{+-}+\ket{-+})$. 

We will use the CNOT gate both in the computational and Hadamard bases. The first qubit of a two-qubit CNOT gate is called the {\em control} qubit, and the second qubit is the {\em target} qubit. The target qubit is flipped when the control qubit is $\ket{1}$, the target qubit remains the same when the control qubit is $\ket{0}$. The CNOT gate in the Hadamard basis has a reverse effect on the control and the target qubits. The control is flipped when the target qubit is $\ket{-}$, and the control remains the same when the target is $\ket{+}$. For example, (assuming the first qubit as control and the second as target) $\ket{+-}=\frac{1}{2}(\ket{0}+\ket{1})(\ket{0}-\ket{1})
=\frac{1}{2}(\ket{00}-\ket{01}+\ket{10}-\ket{11})\xRightarrow{CNOT} \frac{1}{2}(\ket{00}-\ket{01}+\ket{11}-\ket{10})=\ket{--}$. 
Similarly, $\ket{--}=\frac{1}{2}(\ket{0}-\ket{1})(\ket{0}-\ket{1})
=\frac{1}{2}(\ket{00}-\ket{01}-\ket{10}+\ket{11})\xRightarrow{CNOT} \frac{1}{2}(\ket{00}-\ket{01}-\ket{11}+\ket{10})=\ket{+-}$. It is easy to show that $\ket{++}\xRightarrow{CNOT}\ket{++}$, and $\ket{-+}\xRightarrow{CNOT}\ket{-+}$.

\begin{figure}
\begin{subfigure}[b]{.5\textwidth}
\centering
\includegraphics[scale=0.5]{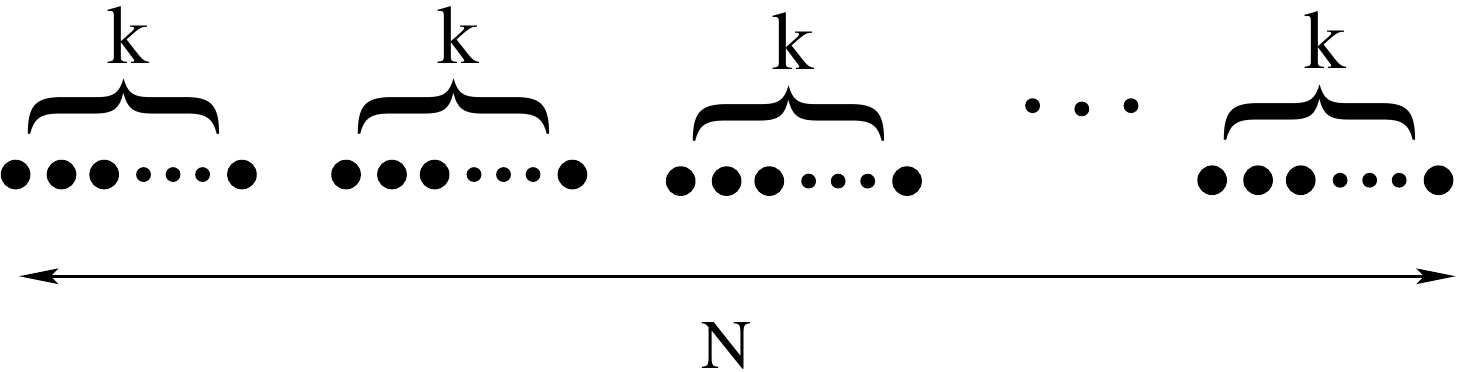}
\caption{Bob needs $Nk$ qubits for receiving $N$ classical bits.}
\label{fig1:first}
\end{subfigure}
\hfill
\begin{subfigure}[b]{.5\textwidth}
\centering
\[\Qcircuit 
{ \lstick{\ket{\psi}} & \ctrl{+1}  & \gate{H} &\meter&\cw&   \\
\lstick{\ket{0}}&\targ &\gate{H}&\meter&\cw 
}\]
\caption{Bob's first decoding circuit.}
\label{fig1:second}
\end{subfigure}
\caption{Illustration for Bob's first decoding algorithm.}
\label{fig1}
\end{figure}

\subsubsection{Bob's first algorithm}

 Alice and Bob share $k$ Bell states $\ket{\Phi^+}$ for each classical bit that Alice wants to send, as shown in Fig.~\ref{fig1:first}. Bob recovers the classical bit with high probability depending on the value of $k$. The circuit is shown in Fig.~\ref{fig1:second}.

For $\ket{\psi_1}$, the result of applying the first CNOT gate is $\ket{00}$. The two Hadamard transforms give the state $\ket{++}$; for $\ket{\psi_2}$, the state after the Hadamard transforms is $\ket{--}$; for $\ket{\psi_3}$, the CNOT operation creates the state $\ket{\Phi^+}$ and the Hadamard gates give the state $\frac{1}{\sqrt 2}(\ket{++}+\ket{--})$. The two measurements are done in the Hadamard basis. Hence, there is no way of distinguishing between these three states through these two measurements, as the measurement results will be either both $\ket{+}$, or both $\ket{-}$. However, if Alice is trying to send the classical bit $1$, she has a probability of $\frac{1}{2}$ to collapse Bob's state to $\ket{\psi_3}$ and probability of $\frac{1}{2}$ to collapse to $\ket{\psi_4}$. If the state collapses to $\ket{\psi_4}$, the CNOT gate produces the state $\ket{\Phi^-}$, and the two Hadamard gates produce the state $\frac{1}{\sqrt 2}(\ket{+-}+\ket{-+})$. Hence the two measurement results will be either $\ket{+}, \ket{-}$, or $\ket{-}, \ket{+}$. This algorithm requires multipe qubits for sending a single classical bit. If Alice wants to send the classical bit $1$, the probability that Bob's state always collapses to $\ket{\psi_3}$ is $\frac{1}{2^k}$ for $k$ qubits. Hence the probability that Bob will always measure $\ket{+},\ket{+}$, or $\ket{-},\ket{-}$ is $\frac{1}{2^k}$ for $k$ qubits. Hence, if Alice uses $k$ qubits to send a single classical bit, the probability that Bob will get the wrong classical bit is $<\frac{1}{2^k}$, which is $<0.001$ for $k=10$. Bob needs $k$ ancilla qubits for decoding each classical bit.

\subsubsection{Bob's second algorithm}

\begin{figure} 
\[\Qcircuit @C=1em @R=.7em 
{ \lstick{\ket{\psi}} & \ctrl{+1}  & \gate{H} &\targ&\qw&\qw&\gate{H}&\ctrl{+1}&\gate{Z}&\ctrl{+3}&\gate{H}&\targ&\qw&\qw  \\ 
\lstick{\ket{0}}&\targ &\gate{H}&\qw&\targ&\qw&\gate{H}&\targ&\qw \\ 
\lstick{\ket{0}} &\qw&\gate{H} &\ctrl{-2}&\ctrl{-1}&\meter&\cw\\
\lstick{\ket{0}} &\qw&\qw&\qw&\qw&\qw&\qw&\qw&\qw&\targ&\gate{H}&\qw&\targ&\qw  \\
\lstick{\ket{0}} &\qw&\qw&\qw&\qw&\qw&\qw&\qw&\qw&\qw&\gate{H}&\ctrl{-4}&\ctrl{-1}&\meter&\cw
}\]
\caption{Bob's second decoding circuit.}
\label{circuit2}
\end{figure}
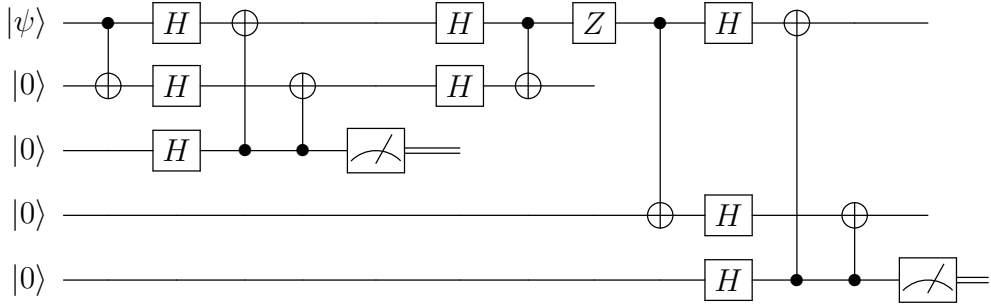

However, Bob can deterministically get the correct classical bit sent by Alice, if he uses a more complex circuit and some ancilla qubits. The circuit is shown in Fig.~\ref{circuit2}. Both the measurements are done in the Hadamard basis. The circuit detects the state $\ket{-}$ in the first measurement if the input state is $\ket{\psi_4}$, hence Bob can detect the classical bit $1$ if Alice's measurement has collapsed Bob's state to $\ket{\psi_4}$. However, the first measurement is inconclusive if the measurement result is $\ket{+}$, all three states $\ket{\psi_1}$, $\ket{\psi_2}$ and $\ket{\psi_3}$ give this result. The rest of the circuit is basically a repeat of the first part after restoring the original state and applying the $Z$ gate to flip the relative  phase. This phase flip has no effect on $\ket{\psi_1}$, introduces a global phase for $\ket{\psi_2}$, which again has no effect. But transforms $\ket{\psi_3}$ to $\ket{\psi_4}$. Hence the result of the second measurement is $\ket{-}$ if the input state is $\ket{\psi_3}$. The measurement results are both $\ket{+}$ if the input state is either $\ket{\psi_1}$ or $\ket{\psi_2}$. These measurement results are summarized in Table~\ref{table1}. We now discuss the details of the circuit in Figure ~\ref{circuit2}. 

\vspace{.3cm}
\noindent
$\ket{\psi_1}$: The first CNOT gate gives the state $\ket{00}$, and the two Hadamard transforms give the state $\ket{++}$. The second ancilla qubit is in the state $\ket{+}$ after the Hadamard transform, and the two CNOT gates with the second ancilla qubit as control do not change the second ancilla qubit from the discussion above. Hence the first measurement result is $\ket{+}$. The two Hadamard gates change the state to $\ket{00}$. The state is $\ket{00}$ after the CNOT gate. The input qubit is restored to $\ket{\psi_1}=\ket{0}$, and the $Z$ gate has no effect. The rest of the circuit is identical to the first part of the circuit, hence the second measurement result is $\ket{+}$. 

\begin{table}[ht]
\centering
\begin{tabular}{ |c| c| c| }
\hline
state & first measurement & second measurement\\ \hline
 $\ket{\psi_1}$ & $\ket{+}$ & $\ket{+}$ \\ 
 $\ket{\psi_2}$ & $\ket{+}$ & $\ket{+}$\\  
 $\ket{\psi_3}$ & $\ket{+}$ & $\ket{-}$\\
 $\ket{\psi_4}$ & $\ket{-}$ & not required\\
 \hline
\end{tabular}
\caption{The measurement results for the second circuit.}
\label{table1}
\end{table}

\vspace{.3cm}
\noindent
$\ket{\psi_2}$: The first CNOT gate gives the state $\ket{11}$, and the two Hadamard gates give the state $\ket{--}$. The second ancilla qubit is $\ket{+}$ after the Hadamard transform. It is flipped by the first qubit to $\ket{-}$ by the first CNOT gate  and flipped back to $\ket{+}$ by the second CNOT gate. Hence the first measurement result is $\ket{+}$. The two Hadamard gates transform the state to $\ket{11}$. The next CNOT gate gives the state $\ket{10}$, and hence the input qubit is restored to $\ket{\psi_2}=\ket{1}$. The $Z$ gate introduces a global phase and the state is $-\ket{1}$. The global phase has no effect on the second part of the circuit and the second measurement result is also $\ket{+}$. 

\vspace{.3cm}
\noindent
$\ket{\psi_3}$: The first CNOT gate gives the entangled state $\frac{1}{\sqrt 2}(\ket{00}+\ket{11}$. The two Hadamard gates give the state $\frac{1}{\sqrt 2}(\ket{++}+\ket{--})$. The first part of the entangled state has no effect on the second ancilla qubit and the second ancilla qubit is flipped twice by the second part and is in the state $\ket{+}$ after the two CNOT gates. Hence the first measurement result is $\ket{+}$. The two Hadamard transforms transform the state to $\frac{1}{\sqrt 2}(\ket{00}+\ket{11})$. The CNOT gate breaks the entanglement and restores the input qubit to $\ket{\psi_3}=\frac{1}{\sqrt 2}(\ket{0}+\ket{1})$. The $Z$ gate transforms the state to $\ket{\psi_4}=\frac{1}{\sqrt 2}(\ket{0}-\ket{1})$, and the effect of the second part of the circuit is identical to the case of $\ket{\psi_4}$ below.

\vspace{.3cm}
\noindent
$\ket{\psi_4}$: The first CNOT gate gives the state $\frac{1}{\sqrt 2}(\ket{00}-\ket{11})$. The Hadamard transforms of the two qubits gives the state $\frac{1}{\sqrt 2}(\ket{+-}+\ket{-+})$. Each term of this state flips the second ancilla qubit $\ket{+}$ exactly once due to the two CNOT gates.  Hence the state of the ancilla qubit after the two CNOT gates is $\ket{-}$, and the first measurement in the Hadamard basis gives $\ket{-}$. There is no need for the second measurement in this case. 

\section{Conclusion}
\label{conclusion}

We have presented two algorithms for communication between spacelike separated locations. The algorithms have different qubit-circuit tradeoffs. Bob's first decoding algorithm requires a simple circuit, but $k$ entangled pairs and $k$ ancilla qubits for transmitting a single classical bit. The second algorithm requires a more complex circuit and four ancilla qubits. Actually the circuit in Figure~\ref{circuit2} can be modified to reuse the second ancilla qubit if the first measurement result is inconclusive and hence three ancilla qubits are sufficient. According to Chefles' theorem~\cite{chefles}, two states can be unambiguously discriminated if and only if they are linearly independent. But the states have to be transformed to an orthogonal set before von Neumann measurement is performed. In the second algorithm, only the state $\frac{1}{\sqrt 2}(\ket{0}-\ket{1})$ transformed to $\frac{1}{\sqrt 2}(\ket{+-}+\ket{-+})$ is orthogonal to both $\ket{++}$ and $\ket{--}$, and hence can be discriminated in the first measurement. The state $\frac{1}{\sqrt 2}(\ket{0}+\ket{1})$ transformed to $\frac{1}{\sqrt 2}(\ket{++}+\ket{--})$ is not orthogonal to $\ket{++}$ and $\ket{--}$, hence the relative phase of $\frac{1}{\sqrt 2}(\ket{0}+\ket{1})$ needs to be flipped and the second measurement is necessary.

One important consequence of our algorithm is that there is no need for classical communication in the {\em quantum teleportation} algorithm~\cite{tele}, if Alice and Bob share extra entangled pairs. Alice can send Bob her two measurement results and Bob can get the two classical bits using Bob's second decoding algorithm in this paper. 





\end{document}